# Geodesy and metrology with a transportable optical clock

Jacopo Grotti[1], Silvio Koller[1], Stefan Vogt[1], Sebastian Häfner[1], Uwe Sterr[1], Christian Lisdat[1], Heiner Denker[2], Christian Voigt[2,3], Ludger Timmen[2], Antoine Rolland[4], Fred N. Baynes[4], Helen S. Margolis[4], Michel Zampaolo[5], Pierre Thoumany[6], Marco Pizzocaro[6], Benjamin Rauf[6,7], Filippo Bregolin[6,7], Anna Tampellini[6,7], Piero Barbieri[6,7], Massimo Zucco[6], Giovanni A. Costanzo[6,7], Cecilia Clivati[6], Filippo Levi[6] and Davide Calonico[6]

[1]Physikalisch-Technische Bundesanstalt, Bundesallee 100, 38116 Braunschweig, Germany
[2]Institut für Erdmessung, Leibniz Universität Hannover, Schneiderberg 50, 30167 Hannover, Germany
[3]GFZ German Research Centre for Geosciences, Telegrafenberg, 14473 Potsdam, Germany
[4]National Physical Laboratory, Teddington, Middlesex TW11 0LW, UK
[5]Laboratoire Souterrain de Modane, Carré Sciences, 1125 Route de Bardonnèche, 73500 Modane, France
[6]Istituto Nazionale di Ricerca Metrologica (INRIM), Physical Metrology Division, Strada delle Cacce 91, 10135 Torino, Italy
[7]Politecnico di Torino, Dipartimento di Elettronica e Telecomunicazioni, Corso duca degli Abruzzi 24, 10125 Torino, Italy

The advent of novel measurement instrumentation can lead to paradigm shifts in scientific research. Optical atomic clocks, due to their unprecedented stability[1,2,3] and uncertainty,[4,5,6,7] are already being used to test physical theories[8,9] and herald a revision of the International System of units (SI).[10,11] However, to unlock their potential for cross-disciplinary applications such as relativistic geodesy,[12] a major challenge remains. This is their transformation from highly specialized instruments restricted to national metrology laboratories into flexible devices deployable in different locations.[13,14,15] Here we report the first field measurement campaign performed with a ubiquitously applicable $^{87}$Sr optical lattice clock.[13] We use it to determine the gravity potential difference between the middle of a mountain and a location 90 km apart, exploiting both local and remote clock comparisons to eliminate potential clock errors. A local comparison with a $^{171}$Yb lattice clock[16] also serves as an important check on the international consistency of independently developed optical clocks. This campaign demonstrates the exciting prospects for transportable optical clocks.

The application of clocks in geodesy fulfils long-standing proposals to interpret a measurement of the fractional relativistic redshift $\Delta\nu_{rel}/\nu_0$ to determine the gravity potential difference $\Delta U = c^2 \Delta\nu_{rel}/\nu_0$ between clocks at two sites (*c* being the speed of light).[12] National geodetic height systems based on classical terrestrial and satellite-based measurements exhibit discrepancies at the decimetre level.[17] Optical clocks, combined with high performance frequency dissemination techniques[18,19] offer an attractive way to resolve these discrepancies, as they combine the advantage of high spectral resolution with small error accumulation over long distances.[18,20]

However, to achieve competitive capability requires high clock performance: a fractional frequency accuracy of $1\times10^{-17}$ corresponds to a resolution of about 10 cm in height. Furthermore, it is important to realize that the side-by-side frequency ratio has to be known to determine the remote frequency shift $\Delta\nu_{rel}$. Taking the uncertainty budgets of optical clocks for granted, harbours the possibility of errors, because very few have been verified experimentally to the low $10^{-17}$ region or beyond.[5,7,18,21] A transportable optical clock not only increases the flexibility in measurement sites but mitigates the risk of undetected errors by enabling local calibrations to be performed.

The test site chosen for our demonstration of chronometric levelling[12] with optical clocks was the Laboratoire Souterrain de Modane (LSM) in France, with the Italian metrology institute INRIM in Torino serving as the reference site. The height difference between the two sites is approximately 1000 m, corresponding to a fractional redshift of about $10^{-13}$. From a geodetic point of view, LSM is a challenging and interesting location in which to perform such measurements: firstly, it is located in the middle of the 13 km long Fréjus road tunnel (rock coverage 1700 m), and secondly the area exhibits long-term land uplift (Alpine orogeny) accompanied by a secular gravity potential variation. Furthermore, LSM lacks the metrological infrastructure and environmental control on which the operation of optical clocks usually relies. The selected location thus constitutes a challenging but realistic testbed with practical relevance.

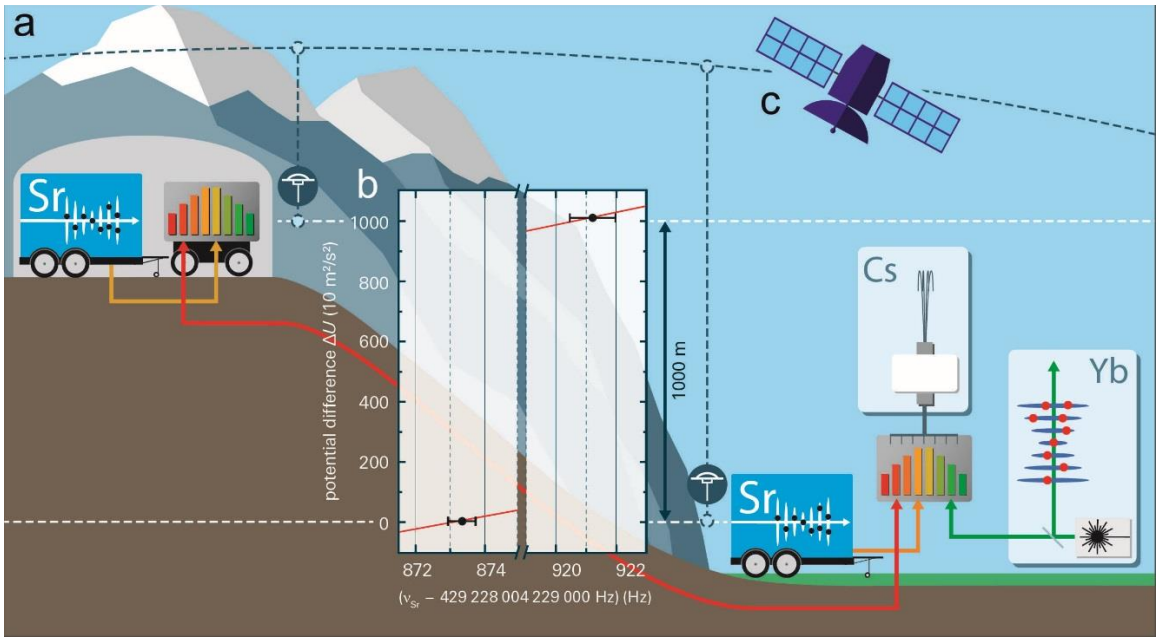

**Fig. 1: Schematic of the measurement campaign**. (**a**) For chronometric levelling the transportable [87]Sr optical lattice clock was placed in the LSM underground laboratory close to the France–Italy border in the Fréjus tunnel (top left). The clock was connected by a noise-compensated fibre link to the Italian national metrology institute INRIM in Torino (red line). There, a primary Cs fountain clock and a [171]Yb optical lattice clock were operated (right). At both sites, frequency combs were used to relate the frequencies of the $^1S_0 - {}^3P_0$ optical clock transitions and the 1.5 µm laser radiation transmitted through the link. After the remote frequency comparison, the transportable clock was moved to INRIM for a side-by-side frequency ratio measurement. (**b**) Frequency of the transportable Sr clock as seen by the INRIM Cs fountain clock (black circles, uncertainties are 1σ). The potential difference $\Delta U$ is based on the geodetic measurement. The red line shows the expected variation of the Sr clock transition frequency due to the relativistic redshift. (**c**) The potential difference between LSM and INRIM was also determined independently by a combination of GNSS, spirit levelling and gravimetric geoid modelling (see Methods).

The transportable [87]Sr lattice clock was operated in both locations, LSM and INRIM, to eliminate the need for *a priori* knowledge of the clock's frequency. The schematic outline of the experiment is given in Fig. 1. LSM and INRIM were connected by a 150 km noise-compensated optical fibre link (see Supplement). At LSM, a transportable frequency comb measured the optical frequency ratio between a laser resonant with the Sr clock transition at 698 nm and 1.542 µm radiation from an ultrastable link laser transmitted from INRIM. In this way, the frequency of the optical clock at LSM could be directly related to the frequency of the link laser even without a highly accurate absolute frequency reference. In addition to the optical carrier, the fibre link was used to disseminate a 100 MHz radio frequency reference signal from INRIM for the frequency comb, frequency counters, and acousto-optic modulators at LSM (see Supplement). At INRIM, a cryogenic Cs fountain clock[22] and a [171]Yb optical lattice clock[16] served as references. The connection between the clocks at INRIM and the link laser is provided by a second frequency comb.

Ten days after arriving at LSM in early February 2016, the first spectra of motional sidebands on the $^1S_0 - {}^3P_0$ clock transition were recorded from the [87]Sr transportable clock. The operation of the lattice clock (see Methods) was similar to the procedure described in previous works. All laser systems required for laser cooling, state preparation and trapping of the Sr atoms in the optical lattice were operated, together with the vacuum system and the control electronics, in an air-conditioned car trailer. The ultrastable interrogation laser for the Sr clock transition was placed next to the trailer in the underground laboratory to avoid its performance being degraded by vibrations induced in the trailer by its air conditioning system. The frequency comb was also operated next to the trailer.

**Tab. 1: Typical uncertainty budgets of the optical clocks.** For the Sr lattice clock, we give the uncertainties for the measurements at LSM (first column) and INRIM (second column).

|  | $^{171}$Yb clock | | $^{87}$Sr clock at LSM | | $^{87}$Sr clock at INRIM | |
| --- | --- | --- | --- | --- | --- | --- |
|  | Correction | Uncertainty | Correction | Uncertainty | Correction | Uncertainty |
| Systematic effect | (in 10$^{-17}$) | | (in 10$^{-17}$) | | (in 10$^{-17}$) | |
| Linear lattice light shift | 4 | 8 | 0 | 24 | 0 | 17 |
| Higher order lattice shifts | 12 | 10 | –1.0 | 0.7 | –0.5 | 0.7 |
| Density shift | 2 | 6 | –1.2 | 3.0 | –2.2 | 5.3 |
| 2$^{nd}$ order Zeeman shift | 27 | 4 | 34.2 | 0.5 | 11.7 | 0.2 |
| BBR | 237.4 | 2.6 | 500.3 | 3.4 | 515.3 | 1.8 |
| Probe light shift | –1.0 | 3.5 | 0.2 | 0.2 | 0.3 | 0.3 |
| dc Stark shift | 0 | 1 | 0 | 0.1 | 0 | 0.1 |
| Servo error | 0 | 1 | 0 | 9.4 | 0 | 3.7 |
| Line pulling | 0 | 0.4 | 0 | 4.1 | 0 | 1.1 |
| Optical path length | 0 | 5 | 0 | 0.8 | 0 | 1.3 |
| AOM switching | 0 | 0.4 | – | – | – | – |
| **Total** | **281** | **16** | **532** | **27** | **524** | **18** |

During this first run of the apparatus in particularly challenging environmental conditions at LSM, the transportable clock operated less reliably than in initial tests before transport.[13] Interruptions were mainly caused by degradation of the light delivery setup for the first cooling stage of the magneto-optical trap (MOT) on the 461 nm $^1S_0 - {}^1P_1$ transition, which was based on a commercial semi-monolithic fibre-based light distribution system. Time-dependent power loss hampered the operation of the Sr lattice clock and thus the evaluation of some contributions to the uncertainty budget of the clock, e.g. the lattice light shift (see Methods), although the blackbody radiation (BBR) shift was still controlled to the level of 3×10$^{−17}$. For these reasons, during the allocated time in the tunnel simultaneous operation was achieved only with the primary Cs fountain clock at INRIM and not with the high stability Yb lattice clock. With the transportable clock operating for 2.8 h over two days at the end of the LSM campaign in mid-March 2016, the instability of the fountain clock (2.2×10$^{−13}\tau^{−1/2}$ where $\tau$ is given in seconds) poses a limitation on the uncertainty of the frequency measurement. The application of a hydrogen maser as a flywheel[23] allowed the measurement time to be extended to 48 h (see Methods) leading to an uncertainty of 17×10$^{−16}$ associated with the limited measurement time. With systematic uncertainties of the Sr and Cs clock of 2.6×10$^{−16}$ and 3×10$^{−16}$ respectively (see Tab. 1, Methods, and Levi *et al.*[22]), the frequency of the Sr lattice clock at LSM was measured by the fountain clock at INRIM with an uncertainty of 18×10$^{−16}$ (see Fig. 1).

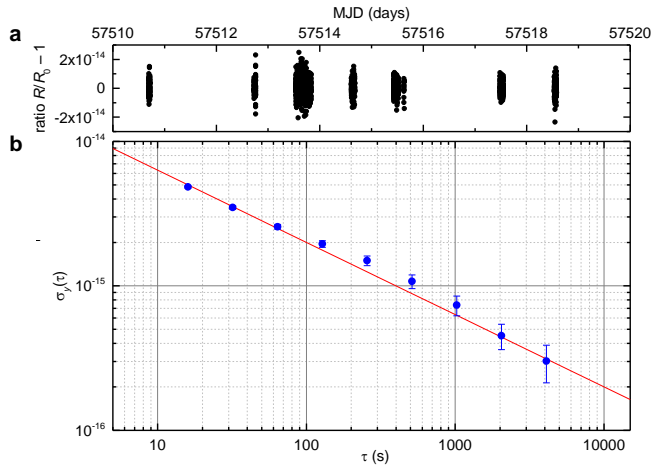

**Fig. 2: Instability of the measured Yb/Sr frequency ratio.** (**a**) Fractional optical frequency ratio $R/R_0$, averaged over 16 s intervals, as a function of the modified Julian date MJD. Here $R_0$ = 1.207 507 039 343 338 122. (**b**) Fractional instability of $R/R_0$ derived from the concatenated data set in (**a**) and expressed as the Allan deviation $\sigma_y$. Error bars denote one standard deviation. The red line depicts an instability of $2\times10^{-14}\,\tau^{-1/2}$.

As discussed above, an independent verification of the experimental observable is required to transform this frequency measurement or clock comparison into a chronometric levelling measurement. For this reason, the Sr apparatus was moved to INRIM in April 2016 for local clock comparisons. There, it was directly linked to the INRIM frequency comb. In the process, small upgrades were made to the setup for the cooling light distribution and the thermal management in the car trailer.

With these changes, the availability of the Sr clock was improved significantly, allowing for several hours of data taking per day after the initial setup phase was completed. With systematic uncertainties comparable to the first campaign and a fountain instability of $3.6\times10^{-13}\,\tau^{-1/2}$, the total uncertainty was reduced by a factor of two to $9\times10^{-16}$ (see Supplement for Sr transitions frequencies). In this chronometric levelling demonstration, we resolved a relativistic redshift of the optical clock lattice clock of 47.92(83) Hz[*] (Fig. 1), from which we infer a potential difference of 10 034(174) $m^2/s^2$. This is in excellent agreement with the value of 10 032.1(16) $m^2/s^2$ determined independently by geodetic means (Methods). Though our result does not yet challenge the classical approach in accuracy, it is a strong first demonstration of chronometric levelling using a transportable optical clock.

With the increased reliability of the transportable Sr clock, we were also able to measure its optical frequency ratio with the Yb lattice clock[16] operated on the $^1S_0 - {}^3P_0$ transition at 578 nm. In total, 31 000 s of common operation of the two optical clocks and the frequency comb were achieved over a period of 7 days. This optical-optical comparison (Fig. 2) shows much higher stability than the optical-microwave one. Consequently, the optical frequency ratio measurement is limited by the systematic uncertainty of the clocks (Tab. 1), rather than by their instability. This demonstrates the key advantage of optical frequency standards: they are able to achieve excellent uncertainties in short averaging times even though they may operate less reliably than their microwave counterparts.

The $^{171}$Yb/$^{87}$Sr frequency ratios measured on different days are summarized in Fig. 3, which also shows previous measurements of this ratio. After averaging (Supplement), we determine the ratio to be $R = \nu_{Yb} / \nu_{Sr} =$ 1.207 507 039 343 338 41(34), which is within two standard deviations of the most accurate previous measurement[24] (Fig. 3). To our knowledge, this is the only optical frequency ratio that has been measured directly by three independent groups.[24,27,28] It therefore constitutes an important contribution to verifying the consistency of optical clocks worldwide.[25] Such measurements are key to establishing more accurate secondary representations of the second[26] as provided by the International Committee for Weights and Measures (CIPM) as a step towards a future redefinition of the SI second.

---

[*] The number in parentheses is the uncertainty referred to the corresponding last digits of the quoted result.

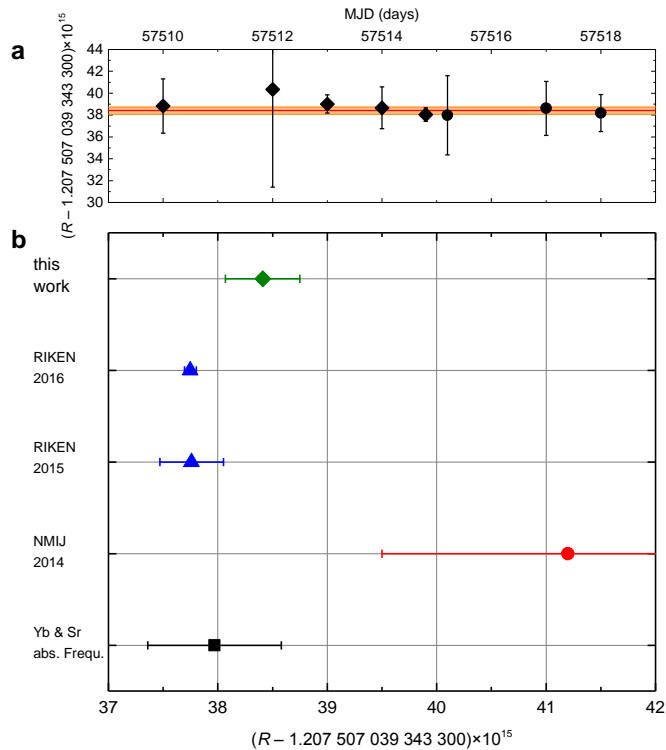

**Fig. 3: Comparison of frequency ratios between $^{171}$Yb and $^{87}$Sr lattice clocks.** (**a**) Averages of the ratios measured on different days (diamonds and circles) and their overall average (line) with its respective uncertainty (colored bar). Diamonds (circles) denote the operation of the Sr lattice clock with a shallow (deep) lattice (see Methods). (**b**) Optical frequency ratios between $^{171}$Yb and $^{87}$Sr have been measured directly in only three groups so far (RIKEN 2016[24] and 2015[27], NMIJ 2014[28] and this work). The lowest point shows the ratio as inferred from averaging all published absolute frequency measurements for $^{171}$Yb and $^{87}$Sr (Supplement).

Note that even with the only slightly improved transportable Sr apparatus as used at INRIM, chronometric levelling against the Yb lattice clock with considerably improved resolution would be possible. We expect that the transportable clock will be able to achieve an uncertainty of $1\times10^{-17}$ or better after a revised evaluation. This uncertainty will enable height differences of 10 cm to be resolved, which is a relevant magnitude for geodesy in regions such as islands, which are hard to access using conventional approaches. As metrological fibre links become more common, chronometric levelling along their paths[29] will become a realistic prospect.

**Acknowledgements**

We like to thank T. Zampieri for his technical support at LSM and A. Mura and Consorzio TOP-IX for technical help in the access to the optical fibre. The authors acknowledge funding from the European Metrology Research Program (EMRP) Project SIB55 ITOC, the EU Innovative Training Network (ITN) Future Atomic Clock Technology (FACT), the DFG funded CRC 1128 geoQ and RTG 1728 and the UK National Measurement System Quantum, Electromagnetics and Time Programme. The EMRP is jointly funded by the EMRP participating countries within EURAMET and the European Union.

**Author contributions**

D.C. coordinated the measurement campaign with contributions from C.L., H.S.M. and Mi.Z.; J.G., S.K., S.V., S.H., U.S. and C.L. designed, built and operated the transportable Sr lattice clock; H.D., C.V. and L.T. performed the geodetic measurements; A.R., F.N.B. and H.S.M. prepared, characterized and operated the transportable frequency comb; M.P., P.T., B.R., F.B., and D.C. designed, built and operated the Yb lattice clock; G.A.C. and F.L. designed, built and operated the INRIM Cs fountain, C.C. and A.T. designed, characterized and operated the optical fibre link between INRIM and LSM; C.C., P.B. and Ma.Z. operated the frequency comb at INRIM. J.G., C.C., M.P., F.L., A.R., F.N.B., H.S.M., S.K. and C.L. contributed to the data analysis for the ratio and absolute frequency measurement. C.L. wrote the paper with support from H.S.M. and D.C. All authors discussed the results and commented on the paper.

**Methods**

### Operation of lattice clocks

The realization and operation of the $^{171}$Yb ($I = 1/2$) and $^{87}$Sr ($I = 9/2$) clocks are very similar and have been presented in detail.[16,13,30] Ytterbium and strontium atoms are cooled to microkelvin temperatures in two-stage magneto-optical traps (MOTs), exploiting the strong $^1S_0 - {}^1P_1$ and weaker $^1S_0 - {}^3P_1$ transitions (at 399 nm and 556 nm for Yb and 461 nm and 689 nm for Sr, respectively). The atoms are then trapped in one-dimensional optical lattices operating at the magic wavelengths[31] $\lambda^{Yb}_{magic} \approx 759$ nm and $\lambda^{Sr}_{magic} \approx 813$ nm.

Finally, the atoms are prepared for spectroscopy in a single magnetic sublevel $m_f$ by optical pumping. As a result, shifts due to cold collisions and line pulling are reduced. The two π-transitions from the $m_f = \pm 1/2$ sublevels in Yb ($m_f = \pm 9/2$ in Sr) are probed alternately at approximately halfwidth detunings so that the interrogation laser is locked to their average transition frequency. This effectively removes the linear Zeeman shift.

### Uncertainties of lattice clocks

Here, we discuss the most important uncertainty contributions listed in Tab. 1. More details are given in references 13, 23 and 16.

*Lattice light shifts:* The lattice of the Yb clock is operated at $\nu_l = 394\,798.238$ GHz with a trap depth $U_0 = 196(4)\,E_r$ ($E_r$ being the lattice recoil energy) and an atomic temperature of 7(3) μK as determined by sideband spectroscopy. We measured the linear shift near the magic wavelength while the nonlinear induced lattice light shift can be calculated using data from Nemitz *et al.*[24]

For the Sr lattice clock, the typical lattice depth was about 100 $E_r$ as measured from sideband spectra. These also yielded an atomic temperature of about 3.5 µK. The light shift cancellation frequency was determined earlier; a reference resonator served as a wavelength reference during the experiments discussed here. The uncertainty of the linear lattice light shift allows for a resonator drift of 50 MHz and changes due to variations of the scalar and tensor light shift.[32] Higher order light shifts were calculated using the coefficients in the same reference. As a check, three of the measurements in Fig. 2 were performed with a deeper lattice of about 160 $E_r$ which resulted in uncertainties for the linear lattice light shift and higher order shifts of $29\times10^{-17}$ and $1\times10^{-17}$, respectively. No significant variation of the measured frequency ratio $R$ was observed.

*Density shift:* The density shift was evaluated in both lattice clocks by varying the interrogated atom number. Corrections for changes of the atomic temperature have been applied for the Sr clock.

*Blackbody radiation (BBR) shift:* The influence of BBR on the clock frequency has been discussed elsewhere.[33,2,34] Temperatures of the atomic environment were measured with calibrated platinum resistance thermometers. The uncertainty of the BBR shift is mostly related to temperature inhomogeneity.

**H-maser as flywheel**

A flywheel oscillator with good stability and high reliability, such as a H-maser, can be used to extend the averaging time between a less reliable system such as our Sr lattice clock and a Cs primary clock with lower stability.[23] The frequency ratio $\nu_{Sr}/\nu_{Cs}$ was thus determined from the frequency ratios $\nu_{Sr}/\nu_H$ and $\nu_H/\nu_{Cs}$ using datasets with different length. The noise of the flywheel means that it had different average frequencies for these two intervals, but the additional uncertainty can be calculated[23] if the noise is well characterized, as it often is for masers. We modelled the maser noise by a superposition of flicker phase noise $6\times10^{-14}\,\tau^{-1}$ ($1\times10^{-13}\,\tau^{-1}$), white frequency noise $5\times10^{-14}\,\tau^{-1/2}$ ($4.5\times10^{-14}\,\tau^{-1/2}$), and a flicker noise $1.7\times10^{-15}$ ($1\times10^{-15}$) in March and (May) 2016, respectively.

**Gravity potential determination**

To provide an accurate reference for the chronometric levelling, we performed a state-of-the-art determination of the gravity (gravitational plus centrifugal) potential with the best possible uncertainty at each clock site. Here, besides the global long-wavelength and eventually the temporal variations[35] of the Earth's gravity potential, the local spatial influence of the gravity potential on the clock frequency needs to be considered. To refine the gravity field modelling around the clock sites and to improve the reliability and uncertainty of the derived geoid model, separate gravity surveys were carried out around INRIM and LSM, resulting in 36 and 123 new gravity points, respectively. The gravity surveys included one absolute gravity observation at INRIM and another at LSM, while the remaining points were observed with relative gravity meters (relative to the established absolute points). Eleven of the gravity points are located inside the Fréjus tunnel near LSM. The new gravity points enabled the existing (largely historic) gravity database to be evaluated (consistency check) and filled some data gaps (coverage improvement).

The determination of the gravity potential involved a combination of Global Navigation Satellite System (GNSS) based ellipsoidal heights at INRIM and near the tunnel portal at LSM, spirit levelling between the GNSS stations and reference points near the clock locations, and a geoid model refined by local gravity measurements.[17] This resulted in a gravity potential difference of 10 029.7(6) m$^2$/s$^2$ between the nearby reference markers, and 10 032.1(16) m$^2$/s$^2$ between the positions of the Sr clock at LSM and INRIM. The assigned uncertainties consider the uncertainty of the geoid model in the Alpine region as well as the uncertainties of the GNSS and levelling measurements; the larger uncertainty for the clocks is due to the simple method used to determine the local height differences between the clocks and the reference markers.

## Supplement:

### Frequency transfer INRIM – LSM

The remote clock comparison was performed by comparing the frequency of a link laser at 1542.14 nm, sent from INRIM to LSM by a telecom optical fibre, to the frequencies of the clocks operated in the two locations. Two fibre frequency combs spanned the spectral gaps between the link laser and the clock interrogation lasers. The combs employed the transfer oscillator principle,[36] making the measurements of the optical frequency ratios immune to the frequency noise of the combs. The frequency of the link laser was stabilized using a high-finesse cavity, whose long term drift is removed by a loose phase-lock to a H-maser via a fibre frequency comb. As a result, the beat notes with the combs remained within a small frequency interval, facilitating long-term operation and reducing potential errors arising from any counter de-synchronization between INRIM and LSM.[18]

The link laser used a multiplexed channel in the telecom fibre. Its path was equipped with two dedicated bidirectional Erbium-doped fibre amplifiers that allowed a phase stable signal to be generated at LSM through the Doppler noise cancellation technique.[18,19] The contribution of fibre frequency transfer to the total fractional uncertainty was assessed to be $3\times10^{-19}$ by looping back the signal from LSM using a parallel fibre. The occasional occurrence of cycle slips was detected by redundant counting of the beat note at INRIM. At LSM, the signal was regenerated by a diode laser phase locked to the incoming radiation with a signal to noise ratio >30 dB in 100 kHz bandwidth; this ensured robust and cycle-slip-free operation.

In addition to the optical reference, a high-quality radio frequency (RF) signal was needed at LSM to operate the Sr clock apparatus (frequency shifters and counters) and the frequency comb. Given the impossibility of having a GNSS-disseminated signal in the underground laboratory, a 100 MHz RF signal was delivered there by amplitude modulation of a second 1.5 μm laser that was transmitted through an optical fibre parallel to the first. At LSM, the amplitude modulation was detected on a fast photodiode, amplified and regenerated by an oven-controlled quartz oscillator (OCXO) at 10 MHz to improve the signal-to-noise ratio. The inherent stability of the free-running fibre link is in this case enough to deliver the RF signal with a long-term instability and uncertainty smaller than $10^{-13}$. This resulting uncertainty contribution to the optical frequency ratio measurement is below $1\times10^{-19}$.

### Averaging of the optical frequency ratio data

We made eight different optical frequency ratio measurements with a total measurement time of 15 h over a period of one week in May 2016 (Fig. 3). The data acquired on different days have different statistical and systematic uncertainties. We applied a statistical analysis that considers the correlations between the measurements coming from the different systematic shifts where the covariance matrix of the eight daily measurements is used to calculate a generalized least squares fit for the average.[25,37] We regarded the systematic uncertainties of the clocks (Tab. 1) as fully correlated, while the statistics related to the measurement duration were uncorrelated.

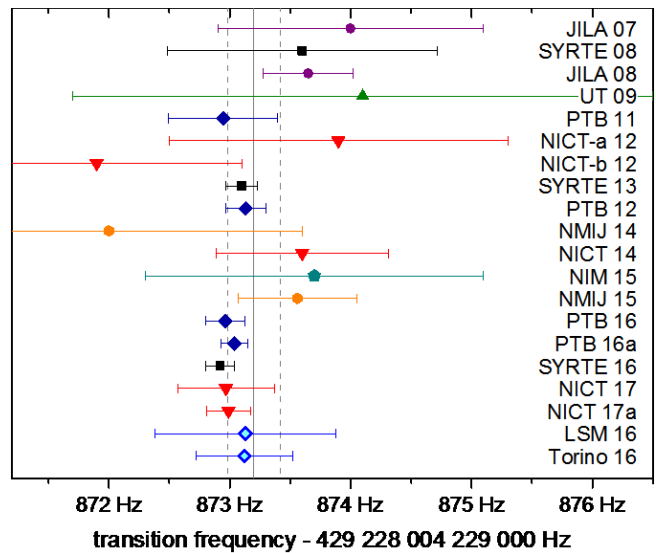

**Fig. Supp1: Comparison of absolute frequency measurements of the Sr clock transition.** The open diamonds at the bottom of the graph show the results from the campaign discussed here. For the LSM data, a correction for the gravitational redshift of – 48.078 Hz as derived from the geodetic data has been applied. The other data have been compiled from various references (38,39,40,41,42,43,44,45,46,47,48,49,50,23,51,52,53). The vertical line indicates the frequency recommended for the secondary representation of the second by Sr lattice clocks[26] and its uncertainty (dashed lines).

**Absolute Frequency of the Sr lattice clock**

The chronometric levelling can be viewed from an alternative perspective: If we assume the conventional measurement of the gravity potential difference is correct then we can deduce an average absolute frequency value of 429 228 004 229 873.13(40) Hz for the Sr lattice clock. The measured frequencies are shown in Fig. Supp1 in comparison with other frequency measurements of the 698 nm clock transition. The accuracy of the absolute frequency measurement achieved with the transportable clock is comparable to several recent measurements with laboratory systems.

Averaging the previously published values shown in Fig. Supp1 gives a frequency of 429 228 004 229 873.05(05) Hz. Together with the average Yb clock transition frequency of 518 295 836 590 863.75(25) Hz derived from other references, [54, 55, 56, 57, 16, 58] a Yb/Sr frequency ratio of 1.207 507 039 343 337 97(61) can be calculated (Fig. 3). This averaging assumes that there are no correlations between the different measurements.